\documentclass[english,prl,aps,longbibliography,floatfix,showpacs,superscriptaddress,twocolumn]{revtex4}
\usepackage[T1]{fontenc}
\usepackage[latin9]{inputenc}
\usepackage{color}
\usepackage{verbatim}
\usepackage{bm}
\usepackage{amsmath}
\usepackage{amssymb}
\usepackage{graphicx}

\makeatletter
\usepackage[colorlinks,linkcolor=blue,anchorcolor=blue,citecolor=blue,urlcolor=blue]{hyperref}

\makeatother

\usepackage{babel}
\begin{document}

\title{Appearance of the universal value $e^{2}/h$ of the zero-bias conductance
in a Weyl semimetal-superconductor junction }

\author{Song-Bo Zhang}

\affiliation{Institute for Theoretical Physics and Astrophysics, University of
Würzburg, D-97074 Würzburg, Germany}

\author{Fabrizio Dolcini}

\affiliation{Dipartimento di Scienza Applicata e Tecnologia del Politecnico di
Torino, I-10129 Torino, Italy }

\author{Daniel Breunig}

\affiliation{Institute for Theoretical Physics and Astrophysics, University of
Würzburg, D-97074 Würzburg, Germany}

\author{Björn Trauzettel}

\affiliation{Institute for Theoretical Physics and Astrophysics, University of
Würzburg, D-97074 Würzburg, Germany}

\date{\today}
\begin{abstract}
We study the differential conductance of a time-reversal symmetric
Weyl semimetal-superconductor (N-S) junction with an s-wave superconducting
state. We find that there exists an extended regime where the zero-bias
differential conductance acquires the universal value $e^{2}/h$ per
unit channel, independent of the pairing and chemical potentials on
each side of the junction, due to a perfect cancellation of Andreev
and normal reflection contributions. This universal conductance can
be attributed to the interplay of the unique spin/orbital-momentum
locking and s-wave pairing that couples Weyl nodes of the same chirality.
We expect that the universal conductance can serve as a robust and
distinct signature for time-reversal symmetric Weyl fermions, and
be observed in the recently discovered time-reversal symmetric Weyl
semimetals.
\end{abstract}
\maketitle

\textit{\textcolor{red}{Introduction.}}\textemdash A Weyl semimetal
(WSM) is a three-dimensional (3D) topological phase of matter in which
the conduction and valence bands touch linearly at discrete points,
called Weyl nodes, in the Brillouin zone near the Fermi energy~\cite{Wan11prb,Burkov11prl,Turner2013arXiv,Hosur13Physique,Armitage2017arXiv170501111A}.
According to the fermion doubling, such Weyl nodes appear in pairs
with opposite chirality \cite{Nielsen81plb,Nielsen81npb}, linked
to monopoles and anti-monopoles of the field of Berry curvature in
momentum space. To ensure nonzero Berry curvature, a WSM must violate
either inversion or time-reversal symmetry. This nontrivial momentum-space
topology of WSMs gives rise to a variety of intriguing physical phenomena,
such as surface Fermi arcs \cite{Wan11prb,Burkov11prl}, the chiral
anomaly \cite{Nielsen83plb,Aji12prbrc,Goswami13prb}, and associated
anomalous transport properties \cite{Fukushima08PRD,Xu11prl,Yang11prb,Zyuzin12prb,Hosur12prl,Son13prb,Burkov14prl,Gorbar14prb,Burkov14prl-chiral,Potter14nc,Lu15Weyl-shortrange,Parameswaran14prx,ZhouJH15prb,ZhangSB16NJP,ZhangCL16nc,LiH16nc,Li16np}.
The actual discoveries of WSMs in a growing number of materials \cite{Halasz12PRB,Hirayama14PRL,LiuJP14prb,Xu15sci-TaAs,Yang15Nphys,Lv15prx,Lv15nphys,Xu15np-NbAs,XuN16NC-TaP,Huang15nc,Weng15prx,Rauch15PRL,Ruan16Ncomms,Ruan16PRL,Jia2016Nmater,Murakamie17scienceAd}
have spurred the interest in investigating the interplay of such topological
phase with other electronic phases and orders.

Recently, possibilities of superconducting states, doping- or proximity-induced,
in WSMs have been discussed \cite{Meng12PRB,Cho12PRB,Shivamoggi13PRB,Yang14PRL,Hosur14PRB,Khanna14PRB,HZWei14PRB,Fischer14prb,Bednik15PRB,BLu15PRL,YLi15Arxiv,Jian15PRL,Liu15PRL,Chen16PRB,WangR16PRB,Bachmanne17scienceAdv}.
Most of the theoretical works investigating hybrid structures based
on WSMs focus on the time-reversal broken case \cite{Uchida14JPSJ,WChen13EPL,Kim16PRB,Khanna16PRB,Madsen17PRB,Bovenzi17PRB,Mukherjee17PRB}.
However, so far almost all the experimentally demonstrated WSMs break
inversion symmetry but preserve time-reversal symmetry \cite{Lv15nphys,Lv15prx,Xu15sci-TaAs,Xu15np-NbAs,Yang15Nphys,XuN16NC-TaP}.
Importantly, while in a time-reversal broken Weyl superconductor the
s-wave pairing couples electrons of opposite chirality, in the time-reversal
symmetric case it couples electrons of the same chirality \cite{Meng12PRB},
so that distinct transport properties in N-S junctions could be expected.

In this work, we study a 3D time-reversal symmetric N-S junction constructed
by a WSM and an s-wave superconducting Weyl metal. Near the Weyl nodes,
the intra-orbital pairing dominates the superconducting state. Denoting
by $\mu_{N}$ and $\mu_{S}$ the chemical potentials of the WSM and
superconductor, respectively, and by $\Delta_{s}$ the superconducting
pairing potential, we find that in the regime $|\mu_{N}|\ll[|\Delta_{s}|^{2}+\mu_{S}^{2}]^{1/2}$,
the contributions of Andreev and normal reflections perfectly cancel
at vanishing excitation energy. In this regime, the zero-bias differential
conductance, thus, takes the universal value $e^{2}/h$ per unit channel,
independent of $\mu_{N}$, $\mu_{S}$, and $\Delta_{s}$. We attribute
this universal conductance to the interplay of the unique spin/orbital-momentum
locking and s-wave pairing in the Weyl junction. We also discuss its
robustness and expect that it can serve as a distinct signature for
time-reversal symmetric Weyl fermions. We are confident that the universal
conductance can be observed in the recently discovered time-reversal
symmetric WSMs \cite{Ruan16Ncomms,Rauch15PRL}.

\textit{\textcolor{red}{Model Hamiltonian.}}\textit{\textemdash }We
start with a low-energy model for a time-reversal symmetric WSM \cite{Note_g_NS}
\begin{align}
\mathcal{H}_{\text{w}}= & \sum_{{\bf {\bf k}}}\psi_{{\bf {\bf k}}}^{\dagger}H({\bf k})\psi_{{\bf {\bf k}}},\label{eq:StartingModel}\\
H({\bf k})= & k_{x}s_{x}\sigma_{z}+k_{y}s_{y}\sigma_{0}+(\kappa_{0}^{2}-|{\bf k}|^{2})gs_{z}\sigma_{0}\nonumber \\
 & +\beta s_{y}\sigma_{y}-\alpha k_{y}s_{x}\sigma_{y},
\end{align}
where ${\bf k}=(k_{x},k_{y},k_{z})$ is the wave vector, the four-component
spinor $\psi_{{\bf k}}=\left(c_{A,\uparrow,{\bf k}},c_{A,\downarrow,{\bf k}},c_{B,\uparrow,{\bf k}},c_{B,\downarrow,{\bf k}}\right)^{T}$
is written in terms of annihilation operators $c_{s,\sigma,{\bf k}}$
with spin indices $\sigma=\uparrow,\downarrow$ and orbital indices
$s=A,B$. Here, $\sigma_{i}$ ($i=0,x,y,z$) are the $2\times2$ identity
and Pauli matrices for the spin-$1/2$ space, and $s_{i}$ ($i=0,x,y,z$)
for the orbital space. $\kappa_{0},\alpha$ and $\beta$ are real
model parameters. The model (\ref{eq:StartingModel}) breaks inversion
symmetry, i.e., $s_{z}H({\bf k})s_{z}\neq H(-{\bf k})$ by the $\beta$
term, but preserves time-reversal symmetry as shown by $\sigma_{y}H^{*}({\bf k})\sigma_{y}=H(-{\bf k})$.
Suppose $0<\beta<\kappa_{0}$, the model (\ref{eq:StartingModel})
has four Weyl nodes at $\pm{\bf Q}_{\pm}$ where ${\bf Q}_{\pm}=\left(\beta,0,\pm k_{0}\right)$
and $k_{0}=[\kappa_{0}^{2}-\beta^{2}]^{1/2}$.

Near the Weyl nodes, we can linearize the model (\ref{eq:StartingModel})
and rewrite it as a sum of four effective Hamiltonians, each describes
the electrons near one of the Weyl nodes %
\begin{align}
 & \mathcal{H}_{\text{w}}=\sum_{\gamma=1,2,3,4}\sum_{{\bf k}}\thinspace'\Psi_{\gamma,{\bf k}}^{\dagger}H_{\gamma}({\bf k})\Psi_{\gamma,{\bf k}},\\
 & \begin{array}{c}
H_{1(2)}({\bf k})=\left(k_{x}\mp\beta\right)\sigma_{x}+k_{y}\sigma_{y}+\left(k_{z}\mp k_{0}\right)\sigma_{z},\\
H_{3(4)}({\bf k})=\left(k_{x}\mp\beta\right)\sigma_{x}+k_{y}\sigma_{y}-\left(k_{z}\pm k_{0}\right)\sigma_{z},
\end{array}\label{eq:subHamiltonians}
\end{align}
where $k_{z}$ has been re-scaled by $1/(2k_{0})$ and $k_{y}$ by
$1/\alpha$, the indices $\gamma=1,2,3,4$ label the Weyl nodes at
${\bf Q}_{+},$$-{\bf Q}_{+},{\bf Q}_{-},-{\bf Q}_{-}$, respectively,
and $\sum_{{\bf k}}'$ indicates that ${\bf k}$ is confined to the
vicinity of the Weyl nodes. The spinors $\Psi_{\gamma,{\bf k}}\equiv\left(\psi_{\gamma,\uparrow,{\bf k}},\psi_{\gamma,\downarrow,{\bf k}}\right)^{T}$
of Weyl nodes are given by $\Psi_{1,{\bf k}}=\Psi_{3,{\bf k}}=[c_{\uparrow,{\bf k}}^{(B)},c_{\downarrow,{\bf k}}^{(A)}]^{T}$
and $\Psi_{2,{\bf k}}=\Psi_{4,{\bf k}}=[c_{\uparrow,{\bf k}}^{(A)},c_{\downarrow,{\bf k}}^{(B)}]^{T}$
with $c_{\uparrow(\downarrow),{\bf k}}^{(s)}=(c_{s,\uparrow,{\bf k}}\pm c_{s,\downarrow,{\bf k}})/\sqrt{2}$.
$H_{1}({\bf k})$ and $H_{2}({\bf k})$ describe the two Weyl nodes
of positive chirality while $H_{3}({\bf k})$ and $H_{4}({\bf k})$
describe the two Weyl nodes of negative chirality. All the Weyl nodes
consist of different orbitals and spins, and exhibit a nontrivial
spin/orbital-momentum locking. They form two time-reversed pairs,
i.e., $\sigma_{y}H_{1}^{*}({\bf k})\sigma_{y}=H_{2}(-{\bf k})$ and
$\sigma_{y}H_{3}^{*}({\bf k})\sigma_{y}=H_{4}(-{\bf k})$, each of
them with definite chirality.

Next, introducing the s-wave superconducting coupling with both intra-
and inter-orbital pairing potentials and projecting onto the spinors
of Weyl nodes, one can see that the inter-orbital pairing is strongly
suppressed due to the mismatch of spins or momenta \cite{Supplementary}.
Suppose the Weyl nodes are well separated and the chemical potential
is close to the Weyl nodes, then only the intra-orbital pairing is
important and reads
\begin{align}
\mathcal{H}_{\text{S}}= & \mathcal{H}_{\text{S}}^{+}+\mathcal{H}_{\text{S}}^{-},\\
\mathcal{H}_{\text{S}}^{+}= & \sum_{{\bf k}}\thinspace'\left[\left(\Delta_{s}c_{1,\uparrow,{\bf k}}^{\dagger}c_{2,\downarrow,-{\bf k}}^{\dagger}+h.c.\right)+\left(1\leftrightarrow2\right)\right],\nonumber \\
\mathcal{H}_{\text{S}}^{-}= & \sum_{{\bf k}}\ '\left[\left(\Delta_{s}c_{3,\uparrow,{\bf k}}^{\dagger}c_{4,\downarrow,-{\bf k}}^{\dagger}+h.c.\right)+\left(3\leftrightarrow4\right)\right].\nonumber
\end{align}
The pairing potential $\Delta_{s}$ couples electrons on Weyl nodes
stemming from the time-reversed pairs. The whole system can thus be
understood as two effectively independent and equivalent subsystems
with opposite chirality. In the following, we will discuss the physics
of the subsystem with positive chirality.

Using the Nambu spinor in real space for positive chirality $\tilde{\Psi}({\bf r})=[c_{1,\uparrow}({\bf r}),c_{1,\downarrow}({\bf r}),c_{2,\uparrow}({\bf r}),c_{2,\downarrow}({\bf r}),c_{1,\downarrow}^{\dagger}({\bf r}),$
$-c_{1,\uparrow}^{\dagger}({\bf r}),c_{2,\downarrow}^{\dagger}({\bf r}),-c_{2,\uparrow}^{\dagger}({\bf r})]^{T}$,
we recast the Hamiltonian in a Bogoliubov-de Gennes (BdG) form
\begin{align}
\mathcal{H}_{+}= & \dfrac{1}{2}\int d{\bf r}\Phi^{\dagger}({\bf r})H_{\text{BdG}}\Phi({\bf r}),\\
H_{\text{BdG}}= & \left(-i{\bf \partial}_{{\bf r}}\cdot{\bf \bm{\sigma}}-\mu\sigma_{0}\right)\tau_{0}\nu_{z}+|\Delta_{s}|e^{i\phi\nu_{z}}\sigma_{0}\tau_{x}\nu_{x},\label{eq:BdG-Hamiltonian}
\end{align}
where $\Delta_{s}({\bf r})=|\Delta_{s}({\bf r})|e^{i\phi({\bf r})}$.
We have introduced the identity and Pauli matrices $\nu_{i}$ and
$\tau_{i}$ ($i=0,x,y,z$) for electron-hole and Weyl-node degrees
of freedom, respectively, and moved the $k_{0}$ and $\beta$ dependence
into the wave function by performing a unitary transformation $\Phi({\bf r})=e^{i\left(k_{0}z\sigma_{z}+\beta x\sigma_{x}\right)\tau_{z}\nu_{z}}\tilde{\Psi}({\bf r})$.
In a uniform system, the eigenenergies are given by $\varepsilon=\pm[|\Delta_{s}|^{2}+(|{\bf k}|\pm\mu)^{2}]^{1/2}$.
The superconductor is fully gapped. The BdG Hamiltonian (\ref{eq:BdG-Hamiltonian})
 decouples into two $4\times4$ identical blocks which can be treated
separately. We will consider one block which is enough to fully describe
the junction problem.

\textit{\textcolor{red}{Reflection probabilities in a Weyl N-S junction.}}\textit{\textemdash }The
time-reversal symmetric Weyl N-S junction can be described by the
BdG Hamiltonian\ (\ref{eq:BdG-Hamiltonian}) with $\Delta_{s}(z)=\Delta e^{i\phi}\Theta(z)$
and $\mu(z)=\mu_{N}\Theta(-z)+\mu_{S}\Theta(z)$. Here $\Theta(z)$
is the Heaviside step function, $\Delta>0$ and a constant superconducting
phase $\phi$ are assumed. The wave vector ${\bf k}_{\parallel}=(k_{x},k_{y})$
parallel to the N-S interface is conserved. We can treat each ${\bf k}_{\parallel}$
separately and work with a quasi-1D junction problem.

Assuming first a clean interface and matching the wave function at
the interface, the probabilities of Andreev and normal reflections
at an excitation energy $\varepsilon\geqslant0$, in general, can
be expressed as
\begin{align}
R_{eh}(\varepsilon,{\bf k}_{\parallel}) & =|\cos(2\alpha_{e})\cos(2\alpha_{h})||\sin(\tilde{\alpha}_{e}-\tilde{\alpha}_{h})/\mathcal{Z}|^{2},\label{eq:ReflectionProbability}\\
R_{ee}(\varepsilon,{\bf k}_{\parallel}) & =\left|\mathcal{Y}/\mathcal{Z}\right|^{2},\label{eq:NormalRflection}
\end{align}
respectively, where $\mathcal{Z}=e^{i\beta}\cos(\alpha_{e}+\tilde{\alpha}_{e})\sin(\alpha_{h}+\tilde{\alpha}_{h})-e^{-i\beta}\cos(\alpha_{e}+\tilde{\alpha}_{h})\sin(\alpha_{h}+\tilde{\alpha}_{e})$,
$\mathcal{Y}=e^{i\beta}\sin(\alpha_{e}-\tilde{\alpha}_{e})\sin(\alpha_{h}+\tilde{\alpha}_{h})-e^{-i\beta}\sin(\alpha_{e}-\tilde{\alpha}_{h})\sin(\alpha_{h}+\tilde{\alpha}_{e})$,
$\alpha_{e(h)}=\arctan(k_{\parallel}/k_{e(h)})/2,$ $\tilde{\alpha}_{e(h)}=\arctan(k_{\parallel}/k_{eq(hq)})/2$
and $k_{\parallel}$$=|{\bf k}_{\parallel}|$. The perpendicular momenta
for the electron (hole) and electronlike (holelike) quasiparticle
are $k_{e(h)}=\text{sgn}(\varepsilon\pm\mu_{N}+k_{\parallel})[(\varepsilon\pm\mu_{N})^{2}-k_{\parallel}^{2}]^{1/2}$
and $k_{eq(hq)}=\text{sgn}\{\varepsilon\pm\text{sgn}(\mu_{S}\pm k_{\parallel})[\Delta^{2}+(\mu_{S}\pm k_{\parallel})^{2}]^{1/2}\}[(\mu_{S}\pm\Omega)^{2}-k_{\parallel}^{2}]^{1/2}$,
respectively. For subgap energies $\varepsilon\leqslant\Delta$, $\Omega=i[\Delta^{2}-\varepsilon^{2}]^{1/2}$
and $\beta=\arccos(\varepsilon/\Delta)$, while for supragap energies
$\varepsilon>\Delta$, $\Omega=\text{sgn}(\varepsilon)[\varepsilon^{2}-\Delta^{2}]^{1/2}$
and $\beta=-i\text{arccosh}(\varepsilon/\Delta)$. Note that $\alpha_{e(h)}$
is always real while $\tilde{\alpha}_{e(h)}$ can be complex. Detailed
derivation is provided in \cite{Supplementary}. For subgap energies,
$\varepsilon\leqslant\Delta$, $R_{eh}+R_{ee}=1$, whereas for supragap
energies, $\varepsilon>\Delta$, $R_{eh}+R_{ee}<1$. For a generic
oblique incidence (${\bf k}_{\parallel}\neq0$) both normal and Andreev
reflection are present, and only for normal incidence (${\bf k}_{\parallel}=0$)
one has perfect Andreev reflection below the gap, since Eqs.\ (\ref{eq:ReflectionProbability})
and (\ref{eq:NormalRflection}) reduce to $R_{eh}=|e^{-2i\beta}|$
and $R_{ee}=0$.

\textit{\textcolor{red}{Differential conductance.}}\textit{\textemdash }At
zero temperature, the differential conductance (per unit area) in
the N-S junction is given by \cite{Blonder82PRB}
\begin{equation}
\dfrac{dI}{dV}\equiv\dfrac{e^{2}}{h}\int\dfrac{d^{2}{\bf k}_{\parallel}}{(2\pi)^{2}}[1-R_{ee}(eV,{\bf k}_{\parallel})+R_{eh}(eV,{\bf k}_{\parallel})],\label{eq:GNS-formula}
\end{equation}
where $eV$ is the bias voltage. Note that only real $k_{e}$ contribute
in Eq.\ (\ref{eq:GNS-formula}). We normalize the conductance to
the value $G_{0}=e^{2}(\mu_{N}+eV)^{2}/(4\pi h)$, corresponding to
the number of available channels at energy $\mu_{N}+eV$ on the N
side. With the expressions\ (\ref{eq:ReflectionProbability}) and
(\ref{eq:NormalRflection}) in Eq.\ (\ref{eq:GNS-formula}), we are
able to analyze the behaviors of the conductance. We concentrate,
in the following, on two particular parameter regimes: (i) $\mu_{S}=\mu_{N}$
($\Delta$ arbitary); and (ii) $\mu_{S}\gg\Delta$ ($\mu_{N}$ arbitary)
\footnote{but $\mu_{N}$ and $\mu_{S}$ are still small enough so that the linearized
model (\ref{eq:subHamiltonians}) is applicable}, which have a distinct zero-bias feature in commom (see below).

\begin{figure}[h]
\centering

\includegraphics[width=7.6cm]{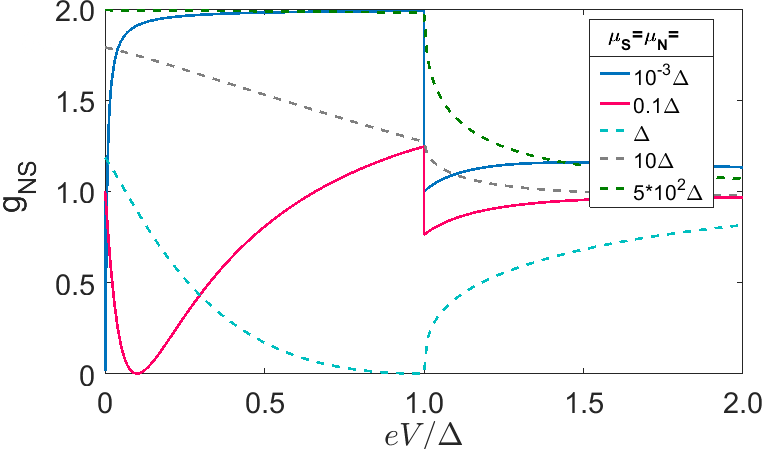}

\caption{Normalized conductance $g_{{\scriptstyle {\scriptscriptstyle \text{NS}}}}$
as a function of bias for various values of $\mu_{N}=\mu_{S}$. }

\label{Fig1:mus=00003Dmuw}
\end{figure}

For regime (i), $\mu_{S}=\mu_{N}$, the normalized conductance $g_{{\scriptscriptstyle \text{NS}}}\equiv G_{0}^{-1}dI/dV$
\footnote{here and everywhere afterwards we refer the normalized conductance
to $g_{{\scriptscriptstyle \text{NS}}}$} as a function of $eV$ is plotted in Fig.\ \ref{Fig1:mus=00003Dmuw}.
At large bias $eV\gg\Delta$, all curves converge to unity. This is
expected since at large excitation energies the influence of superconductivity
is negligible, which together with an identical chemical potential
on both sides makes the interface transparent. The $g_{{\scriptstyle {\scriptscriptstyle \text{NS}}}}$-$eV$
relation is rich in the subgap region, depending on the ratio $\mu_{N}/\Delta$.
For $\mu_{N}/\Delta\gg1,$ the Fermi momentum mismatch of the two
sides is negligible, i.e., $k_{eq(hq)}\approx k_{e(h)}$, thus normal
reflection is suppressed, leading to perfect Andreev reflection with
$g_{{\scriptscriptstyle \text{NS}}}=2$. Similar behavior occurs for
conventional electron systems \cite{Blonder82PRB}. For smaller $\mu_{N}$,
but $\mu_{N}>\Delta$, $g_{{\scriptscriptstyle \text{NS}}}$ bends
down and even shows a dip at $eV=\Delta$. For $0<\mu_{N}<\Delta$,
$g_{{\scriptscriptstyle \text{NS}}}$ vanishes at $eV=\mu_{N}$ as
no hole state is available for Andreev reflection. This is typical
for gapless Dirac systems \cite{Beenakker06PRL}. In the limit $\mu_{N}/\Delta\ll1$,
specular Andreev reflection dominates in the bias region $\mu_{N}<eV<\Delta$
and gives rise to $g_{{\scriptscriptstyle \text{NS}}}=2$ \cite{Supplementary}.
Nevertheless, in the limit of low biases, $g_{{\scriptscriptstyle \text{NS}}}$
approaches unity for $\mu_{N}/\Delta\ll1$ (see solid curves in Fig.\ \ref{Fig1:mus=00003Dmuw}),
implying the universal conductance $e^{2}/h$ per unit channel.

\begin{figure}[h]
\centering

\includegraphics[width=8cm]{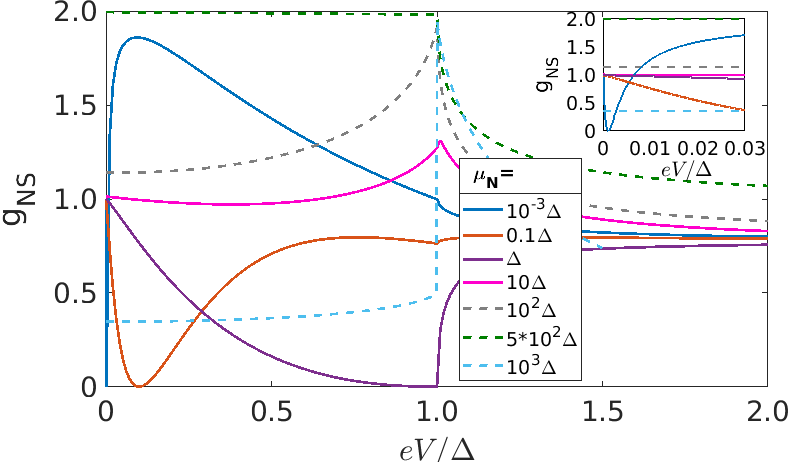}

\caption{Normalized conductance $g_{{\scriptstyle {\scriptscriptstyle \text{NS}}}}$
as a function of bias for fixed $\mu_{S}=5\times10^{2}\Delta$ and
various values of $\mu_{N}$. Inset is the zoom-in in the limit of
low biases. }

\label{Fig1:mus>>Delta}
\end{figure}

Let us now consider regime (ii), $\mu_{S}\gg\Delta$, which corresponds
to the most relevant experimental condition and is depicted in Fig.
\ref{Fig1:mus>>Delta}. For $\mu_{N}>\mu_{S}$, $g_{{\scriptscriptstyle \text{NS}}}$
varies little in the subgap region and it decreases smoothly to a
constant at large bias. With decreasing $\mu_{N}$, $g_{{\scriptscriptstyle \text{NS}}}$
increases in the subgap region or at large bias. For $\mu_{N}=\mu_{S}$,
$g_{{\scriptscriptstyle \text{NS}}}$ is maximized for any bias and
shows perfect Andreev reflection with $g_{{\scriptscriptstyle \text{NS}}}=2$
in the subgap region. For $\mu_{N}<\Delta$, the vanishing of $g_{{\scriptscriptstyle \text{NS}}}$
can also be observed at $eV=\mu_{N}$ where no Andreev reflection
is allowed. Most remarkably, for $\mu_{N}\ll\mu_{S}$, one can notice
again that all the curves approach unity in the limit of low biases,
despite that they vary substantially away from zero bias, and converge
to a constant $4\log(2)-2$ at large bias (see solid curves and inset
in Fig.\ \ref{Fig1:mus>>Delta}).

\begin{figure}[h]
\centering

\includegraphics[width=5cm]{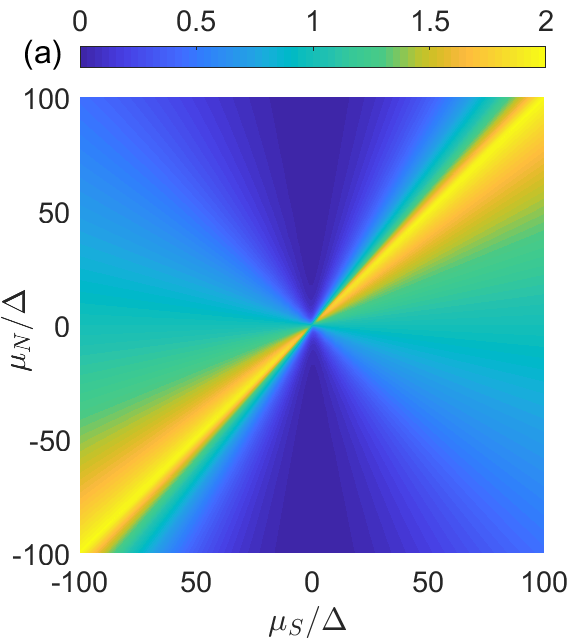}\includegraphics[width=3.5cm]{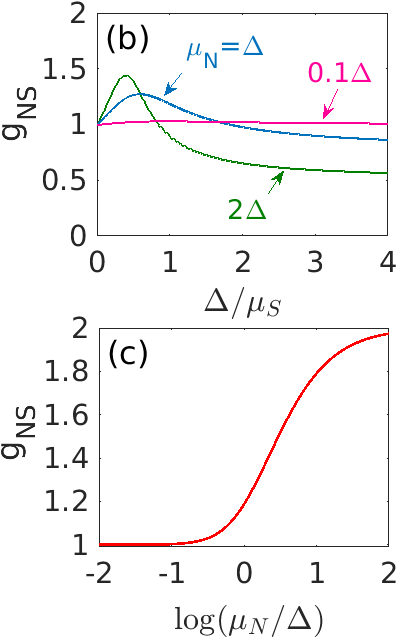}

\caption{(a) The zero-bias conductance $g_{{\scriptscriptstyle \text{NS}}}$
as a function of $\mu_{S}$ and $\mu_{N}$. (b) $g_{{\scriptscriptstyle \text{NS}}}$
as a function of $\Delta/\mu_{S}$ for various values of $\mu_{N}$.
(c) semi-logarithmic plot of $g_{{\scriptscriptstyle \text{NS}}}$
as a function of $\mu_{S}=\mu_{N}$. }

\label{phase-diagram}
\end{figure}

\textit{\textcolor{red}{Zero-bias conductance and universal value.}}\textit{\textemdash }Figure\ \ref{phase-diagram}
focuses on the behavior of the zero-bias conductance $g_{{\scriptscriptstyle \text{NS}}}$.
In particular, Fig.\ \ref{phase-diagram}(a) displays various salient
features of $g_{{\scriptscriptstyle \text{NS}}}$ as a function of
$\mu_{S}$ and $\mu_{N}$. First, $g_{{\scriptscriptstyle \text{NS}}}$
is centrosymmetric in the phase space $\{\mu_{N},\mu_{S}\}$, as a
hallmark of particle-hole symmetry of the system. Second, $g_{{\scriptscriptstyle \text{NS}}}$
shows a ridge along the line $\mu_{N}=\mu_{S}$ where the small Fermi
momentum mismatch strongly suppresses normal reflection. In contrast,
when $|\mu_{S}|\ll|\mu_{N}|$, the Fermi momentum mismatch is large
and normal reflection is enhanced, we have thus vanishing $g_{{\scriptscriptstyle \text{NS}}}$.
Finally, $g_{{\scriptscriptstyle \text{NS}}}$ is always smaller than
unity in the bipolar regime with $\mu_{N}\mu_{S}<0$, implying that
the normal reflection contribution is larger than the Andreev reflection
contribution.

Figure\ \ref{phase-diagram}(b) shows the behavior of $g_{{\scriptscriptstyle \text{NS}}}$
as a function of $\Delta/\mu_{S}$ for various values of $\mu_{N}$.
Figure\ \ref{phase-diagram}(c) instead displays $g_{{\scriptscriptstyle \text{NS}}}$
with respect to $\mu_{N}=\mu_{S}$. The universal conductance $e^{2}/h$
clearly appears in the regime:
\begin{equation}
|\mu_{N}|\ll\sqrt{\Delta^{2}+\mu_{S}^{2}},\label{eq:regime}
\end{equation}
where the Fermi momenta on the two sides of the interface are very
different, i.e, $|k_{e}|\ll|k_{eq}|$. We note that such regime corresponds
to an ideal semimetal phase on the N side, which should be experimentally
accessible. To understand the occurrence of the universal conductance,
we focus on the regime (\ref{eq:regime}) and analyze our analytical
results. Since only real $k_{e}$ contribute to the conductance given
by Eq.\ (\ref{eq:GNS-formula}), the channels with $k_{\parallel}<|\mu_{N}|$
are relevant. From the BdG Hamiltonian\ (\ref{eq:BdG-Hamiltonian}),
we observe that while on the N side the parallel wave vector ${\bf k}_{\parallel}$,
which couples different spins and orbitals, is significant, on the
S side it becomes negligible compared to the perpendicular momentum,
i.e., $k_{\parallel}\ll|k_{eq}|\approx[\Delta^{2}+\mu_{S}^{2}]^{1/2}$.
Thus, the $A$- and $B$-orbital components are decoupled from each
other on the S side. As a result, the reflection probabilities %
at zero energy reduce to
\begin{align}
R_{eh}(0,{\bf k}_{\parallel}) & =1-|k_{\parallel}/\mu_{N}|^{2},\label{eq:SimplifiedAR}\\
R_{ee}(0,{\bf k}_{\parallel}) & =|k_{\parallel}/\mu_{N}|^{2}.\label{eq:SimplifiedNR}
\end{align}
They become functions of a single parameter $|k_{\parallel}/\mu_{N}|$.
Notably,  normal and Andreev reflections have opposite contribution
to the conductance, according to Eq.\ (\ref{eq:GNS-formula}). Plugging
Eqs.\ (\ref{eq:SimplifiedAR}) and (\ref{eq:SimplifiedNR}) into
Eq.\ (\ref{eq:GNS-formula}), it is straightforward to see that the
contributions from Andreev and normal reflections cancel each other
perfectly, giving rise to the universal conductance $e^{2}/h$ per
unit channel. The perfect cancellation in the 3D Weyl junction can
be understood as a result of the unique spin/orbital-momentum locking
and s-wave pairing, which can be inferred from the analog of the Weyl
system to a 1D ferromagnet-superconductor junction \cite{Supplementary}.

\textit{\textcolor{red}{Robustness of the universal value.}}\textit{\textemdash }We
note that in a conventional electron system with parabolic spectrum,
the zero-bias conductance can also exhibit a universal value in the
regime (\ref{eq:regime}). However, it is trivially zero. Indeed,
since in that case velocity and current are linear in momentum, for
large momentum mismatch, the conservation of the flux at the interface
is only possible if the flux vanishes. By contrast, in a Dirac system,
the Fermi velocity is constant and the flux conservation is less sensitive
to the Fermi momentum mismatch. As a consequence, non-vanishing flux
and conductance are possible. In graphene, a 2D Dirac system, a finite
characteristic value ($4e^{2}/3h$) of the zero-bias conductance can
be found \cite{Beenakker06PRL}. However, the instabilities of the
2D Dirac cone to small perturbations, such the intrinsic spin-orbit
coupling \cite{Kane05PRL} or the coupling to the substrate \cite{Giovannetti07PRB},
likely mask such effect. In fact, to the authors' knowledge, the value
$4e^{2}/3h$ in graphene has never been observed experimentally. By
contrast, the Weyl nodes in a WSM are topologically protected and
cannot be gapped out. Therefore, we expect that the universal conductance
$e^{2}/h$ found here is accessible in experiments.

Finally, we stress that the universal conductance predicted by us
is robust in the presence of an interface barrier, due to Klein tunneling
\cite{Katsnelson06nphys,Supplementary}. The interface barrier can
be modeled by a potential term $V_{0}\nu_{z}\Theta(z+d)\Theta(-z)$
in the BdG Hamiltonian, where we assume the barrier length $d\rightarrow0$
and potential $V_{0}\rightarrow\infty$ but the barrier strength $\chi\equiv V_{0}d$
remains finite \cite{Bhattacharjee06PRL}. Then, $g_{{\scriptscriptstyle \text{NS}}}$
is an oscillation function of $\chi$ with a period $\pi$. In the
regime (\ref{eq:regime}), $g_{{\scriptscriptstyle \text{NS}}}$ oscillates
slightly around the universal value, as shown in Fig.\ \ref{phase-diagram}.
\textcolor{black}{Note that if the system is not deep in the regime
(\ref{eq:regime}), only a small deviation from $e^{2}/h$ appears.}
Therefore, the universal conductance can be used as a distinct signature
for time-reversal symmetric Weyl fermions.
\begin{figure}[h]
\centering

\includegraphics[width=8.5cm]{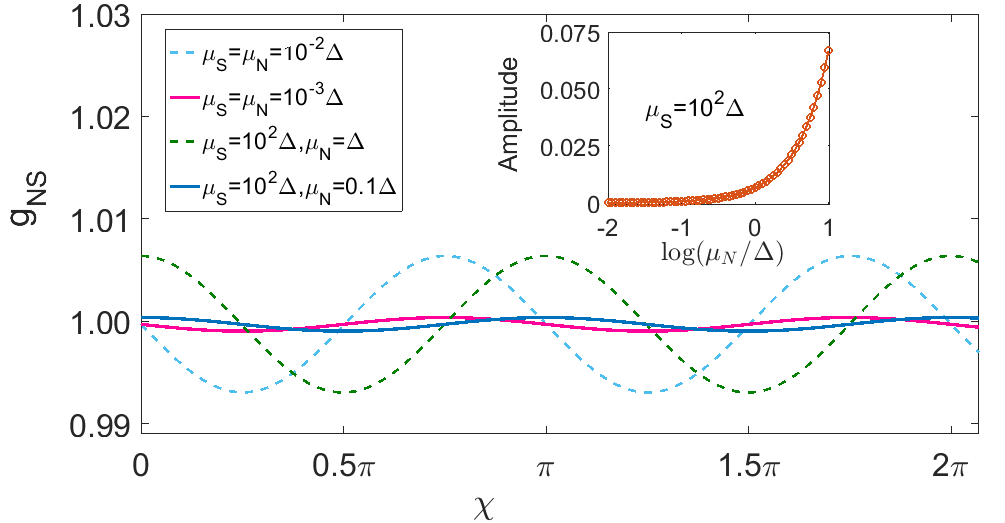}

\caption{The zero-bias conductance $g_{{\scriptscriptstyle \text{NS}}}$ as
a function of the barrier strength $\chi$ for various $\mu_{S}$
and $\mu_{N}$. Inset is the semi-logarithmic plot of the oscillation
amplitudes as a function of $\mu_{N}$ for fixed $\mu_{S}=10^{2}\Delta$.}

\label{Fig:BarrierEffect}
\end{figure}

\textit{\textcolor{red}{Experimental relevance.}}\textit{\textemdash }Recently,
an ideal time-reversal symmetric WSM phase has been proposed in
3D HgTe under compressive strain \cite{Ruan16Ncomms,Rauch15PRL}.
There are likely four pairs of Weyl nodes in the WSM phase \cite{Ruan16Ncomms}.
However, as long as the Fermi energy is close enough to the Weyl nodes,
the system can be decoupled to multiple equivalent time-reversed subsystems.
Then our analysis and main results should hold. Importantly, superconductivity
in 3D compressively strained HgTe could be realized by proximity to
a conventional s-wave superconductor, similar to the case of tensilely
strained HgTe, a 3D topological insulator \cite{Maier12PRL,Wiedenmann17PRB}.
Therefore, we expect that the universal conductance $e^{2}/h$ could
be measured on compressively strained HgTe systems.

\textit{\textcolor{red}{Summary.}}\textit{\textemdash }We have analyzed
a time-reversal symmetric Weyl N-S junction with an s-wave superconducting
pairing state. In an accessible regime, the zero-bias differential
conductance takes the universal value $e^{2}/h$ per unit channel,
independent of the pairing and chemical potentials, as the Andreev
and normal reflection contributions perfectly cancel at vanishing
excitation energy. The universal conductance can be understood as
a consequence of the interplay of the unique spin/orbital-momentum
locking and s-wave pairing in the WSM system.

\textit{\textcolor{red}{Acknowledgments}}\textit{.\textemdash }We
thank Jian Li, Benedikt Scharf, Martin Stehno, and Xianxin Wu for
valuable discussions. This work was supported by by the DFG (SPP1666
and SFB1170 \textquotedbl{}ToCoTronics\textquotedbl{}) and the ENB
Graduate School on \textquotedbl{}Topological Insulators\textquotedbl{}.


%

\newpage
\onecolumngrid
\newpage
\renewcommand{\thesection}{S\arabic{section}}
\renewcommand{\thetable}{S\arabic{table}}
\renewcommand{\thefigure}{S\arabic{figure}}
\numberwithin{equation}{section}
\renewcommand{\thepage}{\arabic{page}}
\setcounter{equation}{0}
\section*{Supplemental material}

In this Supplemental Material, we show (S1) the derivation of the
effective Hamiltonian for the s-wave superconducting pairing; (S2)
Transport probabilities of the Weyl N-S junction; (S3) Analogy of
the Weyl junction to a 1D ferromagnet-superconductor (F-S) junction;
(S4) Effect of an interface barrier on the conductance.

\section{Effective Hamiltonian for the s-wave superconducting coupling}

The s-wave superconducting coupling with both intra- and inter-orbital
pairing potentials is given by
\begin{align}
\mathcal{H}_{\text{S}}=- & \sum_{s=A,B}\sum_{{\bf k}}\left(\Delta_{s}c_{s,\uparrow,{\bf k}}^{\dagger}c_{s,\downarrow,-{\bf k}}^{\dagger}+h.c.\right)-\sum_{{\bf k}}[(\tilde{\Delta}_{s}c_{A,\uparrow,{\bf k}}^{\dagger}c_{B,\downarrow,-{\bf k}}^{\dagger}+h.c.)+\left(A\leftrightarrow B\right)],\label{eq:pairingterm}
\end{align}
where $\Delta_{s}$ and $\tilde{\Delta}_{s}$ measure the amplitudes
of the intra- and inter-orbital pairing potentials. Under the unitary
transformation $c_{\uparrow(\downarrow),{\bf k}}^{(s)}=(c_{s,\uparrow,{\bf k}}\pm c_{s,\downarrow,{\bf k}})/\sqrt{2}$,
$s=A,B$, we have
\begin{equation}
\sum_{{\bf k}}\left(c_{s,\uparrow,{\bf k}}c_{s',\downarrow,-{\bf k}}+c_{s',\uparrow,{\bf k}}c_{s,\downarrow,-{\bf k}}\right)=\sum_{{\bf k}}\left[c_{\downarrow,{\bf k}}^{(s)}c_{\uparrow,{\bf -k}}^{(s')}+c_{\downarrow,{\bf k}}^{(s')}c_{\uparrow,{\bf -k}}^{(s)}\right].
\end{equation}
Thus, we can rewrite Eq.\ (\ref{eq:pairingterm}) in the Nambu spinor
\begin{equation}
\Psi_{{\bf k}}=\left[c_{\uparrow,{\bf k}}^{(A)},c_{\downarrow,{\bf k}}^{(A)},c_{\uparrow,{\bf k}}^{(B)},c_{\downarrow,{\bf k}}^{(B)},c_{\downarrow,-{\bf k}}^{(A)\dagger},-c_{\uparrow,-{\bf k}}^{(A)\dagger},c_{\downarrow,-{\bf k}}^{(B)\dagger},-c_{\uparrow,-{\bf k}}^{(B)\dagger}\right]^{T},
\end{equation}
as
\begin{align}
\mathcal{H}_{\text{S}} & =\dfrac{1}{2}\sum_{{\bf k}}\Psi_{{\bf k}}^{\dagger}H_{\Delta}\Psi_{{\bf k}},\label{eq:A1}
\end{align}
where the BdG Hamitltonian reads
\begin{align}
H_{\Delta} & =\begin{pmatrix}0 & h_{\Delta}\\
h_{\Delta}^{\dagger} & 0
\end{pmatrix},\ \ h_{\Delta}=\begin{pmatrix}\Delta_{s} & 0 & \tilde{\Delta}_{s} & 0\\
0 & \Delta_{s} & 0 & \tilde{\Delta}_{s}\\
\tilde{\Delta}_{s} & 0 & \Delta_{s} & 0\\
0 & \tilde{\Delta}_{s} & 0 & \Delta_{s}
\end{pmatrix}.
\end{align}

At low energy, the whole Nambu spinor containing 16 components in
real space can be written as
\begin{equation}
\Psi({\bf r})=[\Psi_{1,{\bf q}}({\bf r}),\Psi_{2,{\bf q}}({\bf r}),\Psi_{3,{\bf q}}({\bf r}),\Psi_{4,{\bf q}}({\bf r}),\Psi_{1,{\bf q}}^{*}({\bf r}),\Psi_{2,{\bf q}}^{*}({\bf r}),\Psi_{3,{\bf q}}^{*}({\bf r}),\Psi_{4,{\bf q}}^{*}({\bf r})]^{T}.\label{eq:NabmuSpinor}
\end{equation}
where $|{\bf q}|\ll k_{0},\beta$ and the basis functions for the
Weyl nodes read
\begin{align}
\Psi_{1,{\bf q}}({\bf r}) & =\left[\psi_{1,\uparrow}({\bf r}),\psi_{1,\downarrow}({\bf r})\right]=e^{i\beta x+ik_{0}z}e^{i{\bf q}\cdot{\bf r}}\left[c_{\uparrow}^{(B)},c_{\downarrow}^{(A)}\right],\\
\Psi_{2,{\bf q}}({\bf r}) & =\left[\psi_{2,\uparrow}({\bf r}),\psi_{2,\downarrow}({\bf r})\right]=e^{-i\beta x-ik_{0}z}e^{i{\bf q}\cdot{\bf r}}\left[c_{\uparrow}^{(A)},c_{\downarrow}^{(B)}\right],\\
\Psi_{3,{\bf q}}({\bf r}) & =\left[\psi_{3,\uparrow}({\bf r}),\psi_{3,\downarrow}({\bf r})\right]=e^{i\beta x-ik_{0}z}e^{i{\bf q}\cdot{\bf r}}\left[c_{\uparrow}^{(B)},c_{\downarrow}^{(A)}\right],\\
\Psi_{4,{\bf q}}({\bf r}) & =\left[\psi_{4,\uparrow}({\bf r}),\psi_{4,\downarrow}({\bf r})\right]=e^{-i\beta x+ik_{0}z}e^{i{\bf q}\cdot{\bf r}}\left[c_{\uparrow}^{(A)},c_{\downarrow}^{(B)}\right].
\end{align}

The projection of the pairing potential onto the Nambu spinor (\ref{eq:NabmuSpinor})
is calculated as
\begin{equation}
H_{ij}^{\text{S}}=\int d{\bf r}\psi_{i}^{*}({\bf r})H_{S}\psi_{j}({\bf r}),
\end{equation}
where $\psi_{i}$ is the $i$-th component of the Nambu spinor (\ref{eq:NabmuSpinor}).
For illustrations, $H_{1,10}^{\text{S}}$ and $H_{1,12}^{\text{S}}$
are given by
\begin{align}
H_{1,10}^{\text{S}} & =\dfrac{1}{\Omega_{0}}\int d^{3}{\bf r}e^{-i\beta x-ik_{0}z}e^{-i{\bf q}\cdot{\bf r}}\left(0,0,1,0,0,0,0,0\right)H_{\Delta}\left(0,0,0,0,1,0,0,0\right)^{T}e^{-i\beta x-ik_{0}z}e^{i{\bf q}\cdot{\bf r}}\nonumber \\
 & =\dfrac{\tilde{\Delta}_{s}}{\Omega_{0}}\int d^{3}{\bf r}e^{-2i\beta x-2ik_{0}z}=0,
\end{align}
and
\begin{align}
H_{1,12}^{\text{S}} & =\dfrac{1}{\Omega_{0}}\int d^{3}{\bf r}e^{-i\beta x-ik_{0}z}e^{-i{\bf q}\cdot{\bf r}}\left(0,0,1,0,0,0,0,0\right)H_{\Delta}\left(0,0,0,0,0,0,1,0\right)^{T}e^{i\beta x+ik_{0}z}e^{i{\bf q}\cdot{\bf r}}=\Delta_{s}.
\end{align}
Here $\Omega_{0}$ is the volume of the system. In calculating the
element $H_{1,10}^{\text{S}}$, large length $L_{x}$ or $L_{z}$
of the system in the $x$ or $z$ direction or large Weyl-node separation
$k_{0}\ \text{or}\ \beta$ are assumed such that $\beta L_{x}\gg1$
or $k_{0}L_{z}\gg1$ and the integral vanishes. Along these lines,
we obtain the $16\times16$ effective BdG Hamiltonian for the pairing:
\begin{align}
H^{\text{S}} & =\begin{pmatrix}0 & 0 & h_{S} & 0\\
0 & 0 & 0 & h_{S}\\
h_{S}^{\dagger} & 0 & 0 & 0\\
0 & h_{S}^{\dagger} & 0 & 0
\end{pmatrix},\ \ h_{S}=\begin{pmatrix}0 & 0 & 0 & \Delta_{s}\\
0 & 0 & -\Delta_{s} & 0\\
0 & \Delta_{s} & 0 & 0\\
-\Delta_{s} & 0 & 0 & 0
\end{pmatrix}.\label{eq:effectiveBdG}
\end{align}
We can see that the inter-orbital pairing vanishes as $\beta L_{x}\gg1$
or $k_{0}L_{z}\gg1$. To physically understand the vanishing of the
inter-orbital pairing, let us analyze the term $\psi_{\downarrow,{\bf k}}^{(A)}\psi_{\uparrow,{\bf -k}}^{(B)}$.
According to the basis spinors of Weyl nodes (see the main text),
$\psi_{\downarrow,{\bf k}}^{(A)}$ can corresponds to either $\psi_{1,\downarrow,{\bf k}}$
or $\psi_{3,\downarrow,{\bf k}}$, Weyl node 1 or 3. Then the $-{\bf k}$
in $\psi_{\uparrow,{\bf -k}}^{(B)}$ requires that $\psi_{\uparrow,{\bf -k}}^{(B)}$
must correspond to either Weyl node 2 or 4. This coupling, however,
is not allowed since the B-orbital component of Weyl node 2 or 4 always
carries $\downarrow$-spin. Similar analysis can be applied to the
other three terms of the inter-orbital pairing. Therefore, at low
energy the inter-orbital pairing $\tilde{\Delta}_{s}$ is suppressed
and only the intra-orbital pairing $\Delta_{s}$ is important. From
Eq.\ (\ref{eq:effectiveBdG}), we can also observe that $\Delta_{s}$
couples Weyl nodes of the same chirality, i.e., Weyl node 1 to Weyl
node 2 and Weyl node 3 to Weyl node 4. Thus, the whole effective BdG
Hamiltonian decouples into four equivalent $4\times4$ blocks.

\section{Transport probabilities of the N-S junction }

In this section, we apply the Blonder-Tinkham-Klapwijk theory \cite{Blonder82PRB}
to calculate the transport probabilities.

Using one block of the BdG Hamiltonian, the Weyl N-S junction can
be described by
\begin{align}
h_{\text{BdG}} & =\begin{pmatrix}-i\partial_{z}-\mu(z) & k_{x}-ik_{y} & \Delta_{s}(z) & 0\\
k_{x}+ik_{y} & i\partial_{z}-\mu(z) & 0 & \Delta_{s}(z)\\
\Delta_{s}^{*}(z) & 0 & i\partial_{z}+\mu(z) & -k_{x}+ik_{y}\\
0 & \Delta_{s}^{*}(z) & -k_{x}-ik_{y} & -i\partial_{z}+\mu(z)
\end{pmatrix},\label{eq:BdG_nofield}
\end{align}
in the basis $\Phi({\bf r})=[c_{1,\uparrow}({\bf r}),c_{1,\downarrow}({\bf r}),c_{2,\downarrow}^{\dagger}({\bf r}),-c_{2,\uparrow}^{\dagger}({\bf r})]^{T},$
where $\Delta_{s}(z)=\Delta e^{i\phi}\Theta(z)$ and $\mu(z)=\mu_{N}\Theta(-z)+\mu_{S}\Theta(z)$
with $\Delta>0$ and $\Theta(z)$ the Heaviside step function.

On the WSM (N) side, the basis functions for a given excitation energy
$\varepsilon$ can be written as (we neglect the $e^{ik_{x}x+ik_{y}y}$
part for simplicity)
\begin{align}
\varphi_{\overrightarrow{e}}(z) & =(\cos\alpha_{e},e^{i\theta_{{\bf k}}}\sin\alpha_{e},0,0)^{T}e^{ik_{e}z},\label{eq:WF1}\\
\varphi_{\overleftarrow{e}}(z) & =(e^{-i\theta_{{\bf k}}}\sin\alpha_{e},\cos\alpha_{e},0,0)^{T}e^{-ik_{e}z},\\
\varphi_{\overrightarrow{h}}(z) & =(\begin{array}{c}
0,0,-e^{-i\theta_{{\bf k}}}\sin\alpha_{h},\cos\alpha_{h}\end{array})^{T}e^{ik_{h}z},\\
\varphi_{\overleftarrow{h}}(z) & =(0,0,\cos\alpha_{h},-e^{i\theta_{{\bf k}}}\sin\alpha_{h})^{T}e^{-ik_{h}z},\label{eq:WF4}
\end{align}
where $\theta_{{\bf k}}=\arctan(k_{y}/k_{x})$, $\alpha_{e(h)}=\arctan(k_{\parallel}/k_{e(h)})/2$,
and $k_{e(h)}=\text{sgn}(\varepsilon\pm\mu_{N}+k_{\parallel})\sqrt{(\varepsilon\pm\mu_{N})^{2}-k_{\parallel}^{2}}.$
On the superconducting (S) side, the basis functions are
\begin{align}
\varphi_{\overrightarrow{eq}}(z) & =(e^{i\beta}\cos\tilde{\alpha}_{e},e^{i\beta}e^{i\theta_{{\bf k}}}\sin\tilde{\alpha}_{e},e^{-i\phi}\cos\tilde{\alpha}_{e},e^{-i\phi}e^{i\theta_{{\bf k}}}\sin\tilde{\alpha}_{e})^{T}e^{ik_{eq}z},\label{eq:WFS1}\\
\varphi_{\overleftarrow{eq}}(z) & =(e^{i\beta}e^{-i\theta_{{\bf k}}}\sin\tilde{\alpha}_{e},e^{i\beta}\cos\tilde{\alpha}_{e},e^{-i\phi}e^{-i\theta_{{\bf k}}}\sin\tilde{\alpha}_{e},e^{-i\phi}\cos\tilde{\alpha}_{e})^{T}e^{-ik_{eq}z},\\
\varphi_{\overrightarrow{hq}}(z) & =(e^{i\phi}e^{-i\theta_{{\bf k}}}\cos\tilde{\alpha}_{h},e^{i\phi}\sin\tilde{\alpha}_{h},e^{i\beta}e^{-i\theta_{{\bf k}}}\cos\tilde{\alpha}_{h},e^{i\beta}\sin\tilde{\alpha}_{h})^{T}e^{ik_{hq}z},\\
\varphi_{\overleftarrow{hq}}(z) & =(e^{i\phi}\sin\tilde{\alpha}_{h},e^{i\phi}e^{i\theta_{{\bf k}}}\cos\tilde{\alpha}_{h},e^{i\beta}\sin\tilde{\alpha}_{h},e^{i\beta}e^{i\theta_{{\bf k}}}\cos\tilde{\alpha}_{h})^{T}e^{-ik_{hq}z},\label{eq:WFS4}
\end{align}
where $\tilde{\alpha}_{e(h)}=\arctan(k_{\parallel}/k_{eq(hq)})/2,$
and $k_{eq(hq)}=\text{sgn}\left\{ \varepsilon\pm\text{sgn}(\mu_{S}\pm k_{\parallel})\sqrt{\Delta^{2}+(\mu_{S}\pm k_{\parallel})^{2}}\right\} \sqrt{(\mu_{S}\pm\Omega)^{2}-k_{\parallel}^{2}}$.
For subgap energies $\varepsilon\leqslant\Delta$, $\beta=\arccos(\varepsilon/\Delta)$
and $\Omega=i\sqrt{\Delta^{2}-\varepsilon^{2}}$ while, for supragap
energies $\varepsilon>\Delta$, $\beta=-i\text{arccosh}(\varepsilon/\Delta)$
and $\Omega=\text{sgn}(\varepsilon)\sqrt{\varepsilon^{2}-\Delta^{2}}$.
Note that $\alpha_{e(h)}$ is always real while $\tilde{\alpha}_{e(h)}$
can be complex.

At an excitation energy $\varepsilon\geqslant0$, the wave function,
for the scattering state of an electron injected from the WSM and
moving towards the interface, can be described by
\begin{equation}
\Psi(z)=\begin{cases}
\varphi_{\overrightarrow{e}}(z)+b_{0}\varphi_{\overleftarrow{e}}(z)+a_{0}\varphi_{\overleftarrow{h}}(z), & z<0\\
c_{0}\varphi_{\overrightarrow{eq}}(z)+d_{0}\varphi_{\overrightarrow{hq}}(z), & z>0
\end{cases}
\end{equation}
where $a_{0},$ $b_{0},$ $c_{0}$, and $d_{0}$ represent the coefficients
of Andreev and normal reflections, transmissions to two right-moving
quasi-particles, respectively. These coefficients are determined by
the continuity of the wave functions at the N-S interface
\begin{equation}
\Psi(z=0^{+})=\Psi(z=0^{-}).\label{eq:NS-WF-matching}
\end{equation}
With the basis functions and the coefficients, we can calculate the
probabilities of Andreev and normal reflections, and transmissions
which are defined by the Andreev and normal reflected, and transmitted
current densities normalized by the incident current density, respectively.
In general, the transport probabilities can be found, respectively,
as
\begin{align}
R_{eh}(\varepsilon,{\bf k}_{\parallel}) & =\left|\dfrac{\sqrt{\cos(2\alpha_{e})\cos(2\alpha_{h})}\sin(\tilde{\alpha}_{e}-\tilde{\alpha}_{h})}{e^{i\beta}\cos(\alpha_{e}+\tilde{\alpha}_{e})\sin(\alpha_{h}+\tilde{\alpha}_{h})-e^{-i\beta}\cos(\alpha_{e}+\tilde{\alpha}_{h})\sin(\alpha_{h}+\tilde{\alpha}_{e})}\right|^{2},\label{eq:Areflection}\\
R_{ee}(\varepsilon,{\bf k}_{\parallel}) & =\left|\dfrac{e^{i\beta}\sin(\alpha_{e}-\tilde{\alpha}_{e})\sin(\alpha_{h}+\tilde{\alpha}_{h})-e^{-i\beta}\sin(\alpha_{e}-\tilde{\alpha}_{h})\sin(\alpha_{h}+\tilde{\alpha}_{e})}{e^{i\beta}\cos(\alpha_{e}+\tilde{\alpha}_{e})\sin(\alpha_{h}+\tilde{\alpha}_{h})-e^{-i\beta}\cos(\alpha_{e}+\tilde{\alpha}_{h})\sin(\alpha_{h}+\tilde{\alpha}_{e})}\right|^{2},\label{eq:Nreflection}\\
T_{ee}(\varepsilon,{\bf k}_{\parallel}) & =\left|\left|e^{2i\beta}\right|-1\right|\left|\dfrac{\sqrt{\cos(2\alpha_{e})\left(|\cos\tilde{\alpha}_{e}|^{2}-|\sin\tilde{\alpha}_{e}|^{2}\right)}\sin(\alpha_{h}+\tilde{\alpha}_{h})}{e^{i\beta}\cos(\alpha_{e}+\tilde{\alpha}_{e})\sin(\alpha_{h}+\tilde{\alpha}_{h})-e^{-i\beta}\cos(\alpha_{e}+\tilde{\alpha}_{h})\sin(\alpha_{h}+\tilde{\alpha}_{e})}\right|^{2},\label{eq:transmissionPro1}\\
T_{eh}(\varepsilon,{\bf k}_{\parallel}) & =\left|\left|e^{-2i\beta}\right|-1\right|\left|\dfrac{\sqrt{\cos(2\alpha_{e})\left(|\cos\tilde{\alpha}_{h}|^{2}-|\sin\tilde{\alpha}_{h}|^{2}\right)}\sin(\alpha_{h}+\tilde{\alpha}_{e})}{e^{i\beta}\cos(\alpha_{e}+\tilde{\alpha}_{e})\sin(\alpha_{h}+\tilde{\alpha}_{h})-e^{-i\beta}\cos(\alpha_{e}+\tilde{\alpha}_{h})\sin(\alpha_{h}+\tilde{\alpha}_{e})}\right|^{2}.\label{eq:transmissionPro2}
\end{align}
Eqs.\ (\ref{eq:Areflection}) and (\ref{eq:Nreflection}) are the
results {[}Eqs.\ (\textcolor{blue}{8}) and (\textcolor{blue}{9}){]}
given in the main text. In the Dirac system, on requiring the continuity
of the wave function, the continuity of the current flux is also satisfied,
as shown by $R_{ee}+R_{eh}+T_{ee}+T_{eh}=1$. One can see clearly
that for subgap energies $\varepsilon\leqslant\Delta$, $\beta$ is
real, thus there is no transmission probability, i.e., $T_{ee}=T_{eh}=0$.
In the following, we will analyze $R_{eh}$ and $R_{ee}$, since they
are the only functions required in the calculation of the differential
conductance.
\begin{itemize}
\item For normal incidence with ${\bf k}_{\parallel}=0$, $\alpha_{e}=\alpha_{h}=\tilde{\alpha}_{e}=0$
and $\tilde{\alpha}_{h}=\pi/2$. Thus,
\begin{equation}
R_{eh}(\varepsilon,0)=\left|e^{-2i\beta}\right|,\ \ R_{ee}(\varepsilon,0)=0.\label{eq:Areflection-1}
\end{equation}
Andreev reflection dominates, i.e., $R_{eh}=1$, for subgap energies
whereas it decays to zero with increasing $\varepsilon>\Delta$.
\item For $\mu_{N},\mu_{S}\ll\Delta$ and $\mu_{N}<\varepsilon<\Delta$,
$\alpha_{h}=\alpha_{e}$, $\tilde{\alpha}_{h}-\tilde{\alpha}_{e}=\pi/2$,
$\cos(2\alpha_{e})=\sqrt{\varepsilon^{2}-k_{\parallel}^{2}}/\varepsilon,$
$\text{\ensuremath{\sin}(2\ensuremath{\alpha_{e}})=\ensuremath{k_{\parallel}}/\ensuremath{\varepsilon}},$
$\ensuremath{\cos(2\tilde{\alpha}_{e})=\sqrt{\Delta^{2}-\varepsilon^{2}+k_{\parallel}^{2}}/\sqrt{\Delta^{2}-\varepsilon^{2}}}$
and $\sin(2\tilde{\alpha}_{e})=-ik_{\parallel}/\sqrt{\Delta^{2}-\varepsilon^{2}}$.
Thus,
\begin{align}
R_{eh} & (\varepsilon,{\bf k}_{\parallel})=1,\ \ R_{ee}(\varepsilon,{\bf k}_{\parallel})=0.
\end{align}
This indicates that specular Andreev reflection dominates in the region
$\mu_{N}<\varepsilon<\Delta$, leading to $g_{{\scriptscriptstyle \text{NS}}}=2$.
\item At $\varepsilon=\Delta$, $\beta=0$ and $\tilde{\alpha}_{h}-\tilde{\alpha}_{e}=\pi/2$.
Thus,
\begin{align}
R_{eh} & (\Delta,{\bf k}_{\parallel})=\dfrac{\cos(2\alpha_{e})\cos(2\alpha_{h})}{\cos^{2}(\alpha_{e}-\alpha_{h})},\ \ R_{ee}(\Delta,{\bf k}_{\parallel})=\dfrac{\sin^{2}(\alpha_{e}+\alpha_{h})}{\cos^{2}(\alpha_{e}-\alpha_{h})},\label{eq:E=00003DDelta_s}
\end{align}
where the dependence on $\tilde{\alpha}_{e}$ and $\tilde{\alpha}_{h}$
cancels out. The reflection probabilities at excitation energy $\varepsilon=\Delta$
and hence the differential conductance at bias $eV=\Delta$ become
independent of $\mu_{S}$. If $\mu_{N}\ll\Delta$, then $\alpha_{h}=\alpha_{e}$
and
\begin{equation}
R_{eh}(\Delta,{\bf k}_{\parallel})=1-|k_{\parallel}/\Delta|^{2},\ \ R_{ee}(\Delta,{\bf k}_{\parallel})=|k_{\parallel}/\Delta|^{2}.\label{eq:E=00003DDelta}
\end{equation}
Plugging Eqs.\ (\ref{eq:E=00003DDelta}) in Eq.\ (11) in the main
text gives rise to $g_{{\scriptscriptstyle \text{NS}}}=1$. Therefore,
for $\mu_{N},\mu_{S}\ll\Delta$, $g_{{\scriptscriptstyle \text{NS}}}$
shows a jump from 1 to 2 at $eV=\Delta$. \textcolor{black}{For $\mu_{S},\mu_{N}<\Delta$,
the jump at $eV=\Delta$ still appears, but with a smaller discontinuous
value. }
\item At $\varepsilon=\mu_{N}<\Delta$, Eqs.\ (\ref{eq:Areflection}) and
(\ref{eq:Nreflection}) simplify to
\begin{align}
R_{eh}(\mu_{N},{\bf k}_{\parallel})= & 0,\ \ R_{ee}(\mu_{N},{\bf k}_{\parallel})=1.
\end{align}
Andreev reflection is not allowed physically because there is no hole
state on the N side. As a result, the differential conductance vanishes.
The critical energy $\varepsilon=\mu_{N}$ separates two energy regions.
In the region $\varepsilon<\mu_{N}$, Andreev retroreflection occurs
while in the region $\varepsilon>\mu_{N}$, specular Andreev reflection
occurs.
\item At zero energy $\varepsilon=0$, $\beta=\pi/2$, $\alpha_{h}=-\alpha_{e}$,
$\sin\tilde{\alpha}_{h}=\cos^{*}\tilde{\alpha}_{e}$, and $\cos\tilde{\alpha}_{h}=\sin^{*}\tilde{\alpha}_{e}$.
Thus,
\begin{align}
R_{eh}(0,{\bf k}_{\parallel}) & =\left|\dfrac{\cos(2\alpha_{e})(\left|\sin\tilde{\alpha}_{e}\right|^{2}-\left|\cos\tilde{\alpha}_{e}\right|^{2})}{\left|\cos(\alpha_{e}+\tilde{\alpha}_{e})\right|^{2}+\left|\sin(\alpha_{e}-\tilde{\alpha}_{e})\right|^{2}}\right|^{2},\ \ R_{ee}(0,{\bf k}_{\parallel})=\left|\dfrac{2\sin(\alpha_{e}-\tilde{\alpha}_{e})\cos(\alpha_{e}+\tilde{\alpha}_{e})}{\left|\cos(\alpha_{e}+\tilde{\alpha}_{e})\right|^{2}+\left|\sin(\alpha_{e}-\tilde{\alpha}_{e})\right|^{2}}\right|^{2}.
\end{align}
In the regime $|\mu_{N}|\ll\sqrt{\Delta^{2}+\mu_{S}^{2}}$, since
only the channels with real $k_{e}$ and $k_{\parallel}<|\mu_{w}|$
are relevant, we have $k_{\parallel}\ll\left|\mu_{N}\pm\Omega\right|$
and $\tilde{\alpha}_{e}\approx0$. Thus, $R_{eh}$ and $R_{ee}$ further
simplify to
\begin{equation}
R_{eh}(0,{\bf k}_{\parallel})=1-|k_{\parallel}/\mu_{N}|^{2},\ \ R_{ee}(0,{\bf k}_{\parallel})=|k_{\parallel}/\mu_{N}|^{2},\label{eq:universal_probabilities}
\end{equation}
which become functions of a single parameter $k_{\parallel}/\mu_{N}$
and lead to the universal conductance $e^{2}/h$ per unit channel.
\end{itemize}

\section{Analogy of the Weyl junction to a 1D F-S junction }

To see the role played by spin/orbital-momentum locking and s-wave
pairing in the universal conductance $e^{2}/h$, it is instructive
to consider a 1D Dirac F-S junctions. The 1D F-S junction with a ferrromagnet
on the negative side ($z<0$) and a superconductor ($z>0$) on the
positive side can described by
\begin{align}
h_{\text{BdG}}^{(m)} & =\begin{pmatrix}-i\partial_{z}-\mu(z) & m(z) & \Delta(z) & 0\\
m(z) & i\partial_{z}-\mu(z) & 0 & \Delta(z)\\
\Delta(z) & 0 & i\partial_{z}+\mu(z) & -m(z)\\
0 & \Delta(z) & -m(z) & -i\partial_{z}+\mu(z)
\end{pmatrix},\label{eq:BdG_nofield-1}
\end{align}
where $m(z)=m_{0}\Theta(-z)$, $\Delta(z)=\Delta\Theta(z)$ and $\mu(z)=\mu_{F}\Theta(-z)+\mu_{S}\Theta(z)$.
Here the basis is $\Psi=\left(c_{1,\uparrow},c_{1,\downarrow},c_{2,\downarrow}^{\dagger},-c_{2,\uparrow}^{\dagger}\right)^{T}$
with $1$ and $2$ denoting two valleys. Note that the magnetization
$m(z)$ is valley dependent, i.e., it is opposite at the two valleys,
and the pairing potential $\Delta(z)$ couples the same chirality
(defined by the projection of the momentum onto the spin orientation).
This is important to mimic the physics of the Weyl junction.

At zero excitation energy, the basis functions of the right-moving
electron, left-moving electron and left-moving hole on the ferromagnetic
side $z<0$ are given by
\begin{align}
\varphi_{m\overrightarrow{e}}(z) & =(\cos\alpha_{m},\sin\alpha_{m},0,0)^{T}e^{ik_{m}z},\label{eq:N-WF1}\\
\varphi_{m\overleftarrow{e}}(z) & =(\sin\alpha_{m},\cos\alpha_{m},0,0)^{T}e^{-ik_{m}z},\\
\varphi_{m\overleftarrow{h}}(z) & =(0,0,\cos\alpha_{m},\sin\alpha_{m})^{T}e^{ik_{m}z},
\end{align}
respectively, where $\alpha_{m}=\arctan(m_{0}/k_{m})/2$ and $k_{m}=\text{sgn}(\mu_{F})\sqrt{\mu_{F}^{2}-m_{0}^{2}}$.
Note that these zero-energy states on the ferromagnetic side exist
only when $m_{0}<|\mu_{F}|$. Thus, in the following calculation,
we focus on the case of $m_{0}<|\mu_{F}|$. On the S side, the basis
functions of the two ``right-moving'' particles are given by
\begin{align}
\varphi_{m\overrightarrow{eq}}(z) & =\begin{pmatrix}i,0,1,0\end{pmatrix}^{T}e^{i\tilde{k}_{eq}z},\label{eq:S-WF1}\\
\varphi_{m\overrightarrow{hq}}(z) & =\begin{pmatrix}0,1,0,i\end{pmatrix}^{T}e^{-i\tilde{k}_{eq}z},
\end{align}
where $\tilde{k}_{eq}=|\mu_{s}|+i\Delta.$ Both $\varphi_{\overrightarrow{eq}}(z)$
and $\varphi_{\overrightarrow{hq}}(z)$ decay from the interface in
the superconductor as $e^{-z/\xi}$ with $\xi=1/\Delta$.

The matching of the wave function at the interface, $z=0$, gives
rise to the equation
\begin{equation}
\varphi_{m\overrightarrow{e}}(0)+b_{0}\varphi_{m\overleftarrow{e}}(0)+a_{0}\varphi_{m\overleftarrow{h}}(0)=c_{0}\varphi_{m\overrightarrow{eq}}(0)+d_{0}\varphi_{m\overrightarrow{hq}}(0),\label{eq:Equation}
\end{equation}
where $a_{0},b_{0},c_{0}$ and $d_{0}$, similar to the previous section,
represent the coefficients of Andreev reflection, normal reflection
and transmissions, respectively. The coefficients of Andreev and normal
reflections are found as
\begin{align}
a_{0} & =-i\sqrt{1-|m_{0}/\mu_{F}|^{2}},\ \ \ b_{0}=-m_{0}/\mu_{F}.\label{eq:FS-result0}
\end{align}
Then, the probabilities of Andreev and normal reflections are given
by
\begin{equation}
R_{eh}=1-|m_{0}/\mu_{F}|^{2},\ \ R_{ee}=|m_{0}/\mu_{F}|^{2},\label{eq:FS-result}
\end{equation}
respectively. Eq.\ (\ref{eq:FS-result}) indicates that in the absence
of magnetization, $m_{0}=0$, the 1D junction exhibits perfect Andreev
reflection, as expected by the conservation of chirality. A finite
magnetization couples the right and left movers (i.e., different orbitals)
and leads to finite normal reflection. The $\mu_{S}$ and $\Delta$
dependence disappear in the final results\ (\ref{eq:FS-result0})
and (\ref{eq:FS-result}) because the space-dependent phases of the
wave functions drop out in the continuity equation (\ref{eq:Equation}).
Most importantly, one can find that Eqs.\ (\ref{eq:FS-result}) resemble
the form of Eqs.\ (\ref{eq:universal_probabilities}) but with $k_{\parallel}$
replaced by the magnetization $m_{0}$. As a contrast, if the pairing
potential couples opposite chirality (e.g., spin-triplet) or if the
magnetization is valley independent, then following the same approach,
one would find different results.

In the large-momentum mismatch regime $|\mu_{N}|\ll\sqrt{\Delta^{2}+\mu_{S}^{2}}$
of the Weyl N-S junction, the parallel spin/orbital-momentum locking
is significant on the N side but negligible on the S side, thus the
system becomes equivalent to a bundle of 1D Dirac F-S junctions where
the wave vector ${\bf k}_{\parallel}$ acts as the valley-dependent
parallel magnetization. In this way, one can see that the universal
conductance $e^{2}/h$ per unit channel is due to the interplay of
the unique spin/orbital momentum locking and s-wave pairing that couples
Weyl nodes of the same chirality.

\section{Effect of a non-magnetic interface barrier}

In the presence of an interface barrier, the junction can still be
described by the BdG Hamiltonian\ (\ref{eq:BdG_nofield}) but with
\begin{align}
\Delta_{s}(z) & =\Delta e^{i\phi}\Theta(z),\\
\mu(z) & =\mu_{N}\Theta(-z)+\mu_{S}\Theta(z)-V_{0}\Theta(z+d)\Theta(-z).
\end{align}
Here the length $d$ and potential $V_{0}$ of the barrier are assumed
to satisfy
\begin{equation}
d\rightarrow0\ \text{and }V_{0}\rightarrow\infty,\label{eq:BarrierLimit}
\end{equation}
such that the dimensionless barrier strength \textit{$\chi=V_{0}d$
}remain finite \cite{Bhattacharjee06PRL}.

On the N and S sides, the basis functions are still given by Eqs.\ (\ref{eq:WF1}-\ref{eq:WF4})
and (\ref{eq:WFS1}-\ref{eq:WFS4}), respectively. On the barrier,
$0<z<d$, the basis functions can be written as
\begin{align}
\varphi_{\overrightarrow{e}}'(z) & =(1,0,0,0)^{T}e^{-iV_{0}z},\\
\varphi_{\overleftarrow{e}}'(z) & =(0,1,0,0)^{T}e^{iV_{0}z},\\
\varphi_{\overrightarrow{h}}'(z) & =(0,0,0,1)^{T}e^{iV_{0}z},\\
\varphi_{\overleftarrow{h}}'(z) & =(0,0,1,0)^{T}e^{-iV_{0}z}.
\end{align}
Note that these expressions are valid only for the limit (\ref{eq:BarrierLimit}).

For an excitation energy $\varepsilon\geqslant0$, the wave function,
for the scattering state for an electron injected in the WSM and moving
right to the barrier, is described by
\begin{equation}
\Psi(z)=\begin{cases}
\varphi_{\overrightarrow{e}}(z)+a_{0}\varphi_{\overleftarrow{e}}(z)+b_{0}\varphi_{\overleftarrow{h}}(z), & z<-d\\
A\varphi_{\overrightarrow{e}}'(z)+B\varphi_{\overleftarrow{e}}'(z)+C\varphi_{\overleftarrow{h}}'(z)+D\varphi_{\overrightarrow{h}}'(z), & -d<z<0\\
c_{0}\varphi_{\overrightarrow{eq}}(z)+d_{0}\varphi_{\overrightarrow{hq}}(z), & z>0
\end{cases}
\end{equation}
The coefficients, $a_{0},b_{0},c_{0},d_{0},A,B,C,$ and $D$ are found
by matching the wave function at the two interfaces $z=-d$ and $z=0$.
Then, the Andreev and normal reflection probabilities are obtained,
respectively, as
\begin{align}
R_{eh}(\varepsilon,{\bf k}_{\parallel}) & =|\cos(2\alpha_{e})\cos(2\alpha_{h})|\left|\sin(\tilde{\alpha}_{e}-\text{\ensuremath{\tilde{\alpha}}}_{h})/\mathcal{Z}'\right|^{2},\label{eq:ProReh-barrier1}\\
R_{ee}(\varepsilon,{\bf k}_{\parallel}) & =\left|\mathcal{Y}'/\mathcal{Z}'\right|{}^{2},\label{eq:ProReh-barrier2}
\end{align}
where
\begin{align}
\mathcal{Z}'= & e^{i\beta}\left(e^{-i\chi}\sin\alpha_{e}\sin\text{\ensuremath{\tilde{\alpha}}}_{e}-e^{i\chi}\cos\alpha_{e}\cos\text{\ensuremath{\tilde{\alpha}}}_{e}\right)\left(e^{i\chi}\sin\alpha_{h}\cos\text{\ensuremath{\tilde{\alpha}}}_{h}+e^{-i\chi}\cos\alpha_{h}\sin\text{\ensuremath{\tilde{\alpha}}}_{h}\right)\nonumber \\
 & -e^{-i\beta}\left(e^{-i\chi}\sin\alpha_{e}\sin\text{\ensuremath{\tilde{\alpha}}}_{h}-e^{i\chi}\cos\alpha_{e}\cos\text{\ensuremath{\tilde{\alpha}}}_{h}\right)\left(e^{i\chi}\sin\alpha_{h}\cos\text{\ensuremath{\tilde{\alpha}}}_{e}+e^{-i\chi}\cos\alpha_{h}\sin\text{\ensuremath{\tilde{\alpha}}}_{e}\right),\\
\mathcal{Y}'= & e^{i\beta}\left(e^{-i\chi}\cos\alpha_{e}\sin\text{\ensuremath{\tilde{\alpha}}}_{e}-e^{i\chi}\sin\alpha_{e}\cos\text{\ensuremath{\tilde{\alpha}}}_{e}\right)\left(e^{i\chi}\sin\alpha_{h}\cos\text{\ensuremath{\tilde{\alpha}}}_{h}+e^{-i\chi}\cos\alpha_{h}\sin\text{\ensuremath{\tilde{\alpha}}}_{h}\right)\nonumber \\
 & -e^{-i\beta}\left(e^{-i\chi}\cos\alpha_{e}\sin\text{\ensuremath{\tilde{\alpha}}}_{h}-e^{i\chi}\sin\alpha_{e}\cos\text{\ensuremath{\tilde{\alpha}}}_{h}\right)\left(e^{i\chi}\sin\alpha_{h}\cos\text{\ensuremath{\tilde{\alpha}}}_{e}+e^{-i\chi}\cos\alpha_{h}\sin\text{\ensuremath{\tilde{\alpha}}}_{e}\right).
\end{align}
We can see that $R_{eh}$ and $R_{ee}$ are periodic functions of
$\chi$ with a period $\pi$. Thus, the differential conductance is
also a periodic function of $\chi$. Under the condition $\chi=N\pi,\ N=0,\pm1,\pm2,\cdots,$
the expressions (\ref{eq:ProReh-barrier1}) and (\ref{eq:ProReh-barrier2})
reproduce the results Eqs.\ (\ref{eq:Areflection}) and (\ref{eq:Nreflection})
in the absence of the barrier.

At zero excitation energy $\varepsilon=0$, $\beta=\pi/2$, $\alpha_{h}=-\alpha_{e}$,
$\sin\tilde{\alpha}_{h}=\cos^{*}\tilde{\alpha}_{e}$ and $\cos\tilde{\alpha}_{h}=\sin^{*}\tilde{\alpha}_{e}$.
Thus,
\begin{align}
R_{eh}(0,{\bf k}_{\parallel}) & =\left|\dfrac{\cos(2\alpha_{e})\left(|\sin\tilde{\alpha}_{e}|^{2}-|\cos\tilde{\alpha}_{e}|^{2}\right)}{|e^{-i\chi}\sin\alpha_{e}\sin\text{\ensuremath{\tilde{\alpha}}}_{e}-e^{i\chi}\cos\alpha_{e}\cos\text{\ensuremath{\tilde{\alpha}}}_{e}|^{2}+|e^{i\chi}\sin\alpha_{e}\cos\text{\ensuremath{\tilde{\alpha}}}_{e}-e^{-i\chi}\cos\alpha_{e}\sin\text{\ensuremath{\tilde{\alpha}}}_{e}|^{2}}\right|^{2},\label{eq:ProReh-barrier1-1}\\
R_{ee}(0,{\bf k}_{\parallel}) & =\left|\dfrac{\left(e^{i\chi}\sin\alpha_{e}\cos\text{\ensuremath{\tilde{\alpha}}}_{e}-e^{-i\chi}\cos\alpha_{e}\sin\text{\ensuremath{\tilde{\alpha}}}_{e}\right)\left(e^{i\chi}\cos\alpha_{e}\cos\tilde{\alpha}_{e}-e^{-i\chi}\sin\alpha_{e}\sin\tilde{\alpha}_{e}\right)}{|e^{-i\chi}\sin\alpha_{e}\sin\text{\ensuremath{\tilde{\alpha}}}_{e}-e^{i\chi}\cos\alpha_{e}\cos\text{\ensuremath{\tilde{\alpha}}}_{e}|^{2}+|e^{i\chi}\sin\alpha_{e}\cos\text{\ensuremath{\tilde{\alpha}}}_{e}-e^{-i\chi}\cos\alpha_{e}\sin\text{\ensuremath{\tilde{\alpha}}}_{e}|^{2}}\right|{}^{2}.\label{eq:ProReh-barrier2-1}
\end{align}
We now focus on the regime $|\mu_{N}|\ll\sqrt{\Delta^{2}+\mu_{S}^{2}}$.
Since we are considering the channels with real $k_{e}$ and $k_{\parallel}<|\mu_{w}|$,
we have $k_{\parallel}\ll\left|\mu_{N}\pm\Omega\right|$ and $\tilde{\alpha}_{e}\approx0$.
$R_{eh}$ and $R_{ee}$ further simplify to
\begin{equation}
R_{eh}(0,{\bf k}_{\parallel})=1-|k_{\parallel}/\mu_{N}|^{2},\ \ R_{ee}(0,{\bf k}_{\parallel})=|k_{\parallel}/\mu_{N}|^{2},
\end{equation}
which are the same results as those in the absence of the barrier.
The barrier becomes effectively transparent in the regime\textit{
}$|\mu_{N}|\ll\sqrt{\Delta^{2}+\mu_{S}^{2}}$\textit{.} As a result,
the contributions of Andreev and normal reflections cancel perfectly
and the zero-bias differential conductance can still acquire the universal
value $e^{2}/h$ per unit channel.%

\end{document}